\documentclass[12pt,reqno]{amsart}
\usepackage{graphicx}
\usepackage{amssymb,amscd,amsbsy}
\usepackage{amsthm}
\newcommand{\lb}{\linebreak}
\newcommand{\noi}{\noindent}

\newcommand{\PP}{{\mathcal P}}

\newcommand{\RB}{{\mathbb R}}

\newcommand\Dg{\mathfrak D}
\newcommand{\Rg}{{\mathfrak R}}

\newcommand{\HH}{{\mathcal H}}
\newcommand{\MM}{{\mathcal M}}

\def\W{{\mathcal W}}

\def\U{\Upsilon}

\newcommand{\X}{\mathcal X}

\def\c{\omega}
\def\ct{\tilde{\omega}}
\def\[{\left[}
\def\]{\right]}
\def\({\left(}
\def\){\right)}

\def\l{\lambda}
\def\lb{\overline{\l}}

\def\m{\mu}
\def\lt{\frac{\lambda}{2}}
\def\psib{\overline{\psi}}

\def\d{\delta}

\def\ea{\epsilon_a}
\def\epm{\epsilon_{\pm}}

\def\U{\Upsilon}

\def\mi{\mu_{\infty}(\lambda)}
\def\pin{p_{\infty} (\lambda)}
\def\fr{\f_{\rightarrow}}
\def\fl{\f_{\leftarrow}}
\def\Gin{\Gamma_\infty}

\def\G{\Gamma}

\def\12{{1\over 2}}

\newcommand{\db}{{\boldsymbol \delta}}
\newcommand{\e}{{\boldsymbol e}}

\newcommand{\f}{{\boldsymbol f}}

\newcommand{\LLL}{{\boldsymbol L}}

\newcommand{\oneb}{\mathbf 1}
\newcommand{\gb}{{\boldsymbol g}}
\newcommand{\gs}{{\mathfrak g}}
\newcommand{\jo}{{\boldsymbol j}}
\newcommand{\js}{{\mathfrak j}}
\newcommand{\g}{\gamma}

\newcommand{\xit}{\tilde{\xi}}
\newcommand{\bb}{\bar{b}}

\newcommand{\eeq}{\end{equation}}
\newcommand{\beq}{\begin{equation}}
\newcommand{\bay}{\begin{eqnarray}}
\newcommand{\ey}{\end{eqnarray}}
\newcommand{\bey}{\begin{eqnarray*}}
\newcommand{\eey}{\end{eqnarray*}}

\newcommand{\ph}{\operatorname{ phase}}

\newcommand{\sign}{\operatorname{sign}}

\newtheorem{thm}{\hspace{\parindent}Theorem}[section]

\newtheorem{lem}[thm]{\hspace{\parindent}Lemma}

\theoremstyle{remark}
\newtheorem{rem}[thm]{Remark}
\newtheorem*{rem*}{Remark}

\begin{document}

\newcommand{\vse}{\vspace{.2in}}
\numberwithin{equation}{section}

\title{\bf The Atiyah--Hitchin bracket  for  the cubic Nonlinear Schr\"{o}dinger equation. IV.  \newline the  
  scattering potentials.}
\author{K.L.  Vaninsky}
\thanks{ The work is partially supported by NSF grant DMS-9971834. The author thanks Max Plank Institute, Bonn,  and  IHES, Bures sur Ivette,  for the excellent  conditions during his stay at these institutions.}
\begin{abstract}
This is the last  in a series of four  papers on Poisson formalism  for the cubic nonlinear Schr\"{o}dinger 
equation with repulsive nonlinearity.  In this paper we consider scattering   potentials.   
\end{abstract}

\maketitle
\setcounter{section}{0}
\setcounter{equation}{0}
\section{ Introduction}

\subsection{ Statement of the problem.} 
 
In this paper we consider the cubic NLS with repulsive nonlinearity\footnote{Prime $'$  signifies the derivative in the   variable $x$ and dot $\bullet$ the derivative with 
respect to time.}  
$$
i \psi^\bullet= - \psi'' + 2 |\psi |^2 \psi,
$$
where $\psi=\psi(x,t)$ is a complex decaying function  on the entire line, {\it i.e.} $x \in \RB^1$. 

The standard assumption on the decaying potential is that it is summable ($\psi \in L^1(\RB^1$)). 
Such  potentials are called scattering potentials. 
In this paper we assume that the phase space $\MM$ consists of all    Schwartz functions. This would imply, in particular,  an existence of the infinite series of  integrals of motion, {\it etc}. 
Most of our  considerations without difficulty can be extended to the case of summable functions. 

The cubic  Schr\"{o}dinger  equation  is  a Hamiltonian system 
$$
\psi^{\bullet}= \{\psi, \HH\},
$$
with  the bracket
\beq\label{cb}
\{A,B\}=2i \int\limits_{\RB^1} {\d A\over \d \psib(x)}{\d B\over
\d \psi(x)}- {\d A\over \d \psi(x)}{\d B\over 
\d \psib(x)}\, dx; 
\eeq
and   Hamiltonian 
$$\HH={1\over 2} \int\limits_{\RB^1}  |\psi'|^2 
+|\psi|^4 \, dx. 
$$ 

The  equation arises as a compatibility condition for  the commutator relation for some specially chosen differential operators. This  leads to an auxiliary linear spectral problem for the  Dirac operator
\beq\label{dirac}
 \Dg \f= \[\(\begin{array}{ccccc} 1& 0 \\ 0 &  -1  \end{array}\) i\partial_x +
\(\begin{array}{ccccc}   0& -i \psib \\
 i \psi & 0 \end{array}\)\] \f= \lt \f,
\eeq
acting in the space of vector-functions $  \f^T=( f_1, f_2)$.

The relation of the Hamiltonian formalism and complex geometry of the  Dirac operator with general continuous potentials was considered by us in \cite{V3}. 
The Dirac  spectral problem for  rapidly decaying potentials  
can be treated by the methods of scattering theory. 
In this paper we extend our general  approach  introduced in  \cite{V3}  to these scattering potentials and  exploit some specific features 
that arise in this case.

\subsection{Description of results.}  
In their seminal paper devoted to the periodic KdV problem Dubrovin and Novikov \cite{DN}  wrote

\noi
{\it We would like to emphasize  the difficulty of construction of angle type variables in the periodic case in comparison with the scattering case. }
\vskip .1in
\noi

The  situation is similar  for the nonlinear Schr\"{o}dinger equation. Two cases, the periodic and the scattering substantially vary in  difficulty.  
The construction of the action--angle variables for the scattering potentials of NLS equation is known for more then thirty 
years since the work of Zakharov and Manakov, \cite{ZM}. 
The action--angle variable for the periodic NLS problem were considered in \cite{MV}. They are constructed by introducing the Dirac spectral curve and 
the divisor on it. The flows are linearized on the (extended) Jacobian by using a version of the Abel map.  

This   paper  eliminates the  differences between the two cases.  
We utilize the previous approaches of Venakides, \cite{Ven},  and Ercolani and McKean, \cite{EM}.   
The analogs of  spectral curve, divisor and the Abel map are introduced and studied   now in the scattering situation. 

\begin{figure}[htb]
\includegraphics[width=0.60\textwidth]{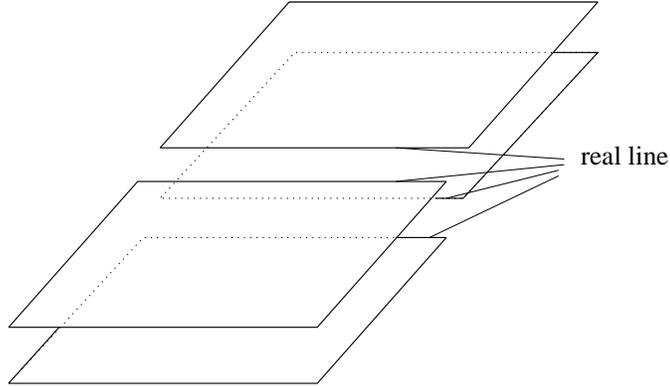}
\caption{ The spectral cover.}
\end{figure} 

\noi

Specifically, we consider  $\G$   a two sheeted 
covering of the complex plane of the  spectral parameter  cut along the real line. 
We introduce the meromorphic function $\Pi(x,Q)$ on the spectral cover $\G$. 
With the help of this function we construct the 
so--called scattering divisor. Then using the scattering divisor we construct  the continuous analog of the Abel map which linearizes the flow. 
Finally, we compute a closed form of the Poisson bracket  for the function $\Pi(x,Q)$. 

\subsection{Thanks.}
We conclude the introduction expressing thanks to  A. Its, H. McKean and I. Krichever for
stimulating discussions. 

\newpage
\section{The Spectral Problem}

\subsection{The NLS hierarchy.} We consider the 
NLS equation 
\bay\label{nls}
i \psi^{\bullet}= -\psi'' + 2 |\psi |^2 \psi, 
\ey
on the line, {\it i.e.}  $x\in \RB^1$.  
We assume that the function $\psi=\psi(x,t)$ belongs to  the Schwartz'
space $S(\RB^1)$ of complex rapidly decreasing infinitely differentiable functions  such that
$
\sup_{x} |(1+x^2)^n \, \psi^{(m)}(x)|\, < \infty, \quad m,\,n=0,1,\ldots.
$ 
 
The NLS is  a Hamiltonian system
$
\psi^{\bullet}= \{\psi, \HH\},
$
with  Hamiltonian
$\HH=\frac{1}{ 2} \int\limits_{\RB^1}  |\psi'|^2 +|\psi|^4 \, dx=${\it energy} and the bracket
\beq\label{nlspb}
\{A,B\}=2i \int\limits_{\RB^1}  \frac{\d A}{ \d \psib(x)}\frac{\d B}{ \d
\psi(x)}- \frac{\d A}{ \d \psi(x)}\frac{\d B}{ \d \psib(x)}\, dx.
\eeq

The NLS equation   is a compatibility condition for the zero curvature relation
$
[\partial_t- V_3,  \partial_x-V_2]=0, 
$
with\footnote{
Here and below  $\sigma$  denotes the {\it Pauli matrices}
$$
\sigma_1=\left(\begin{array}{ccccc}
 0& 1\\
1 & 0
\end{array}  \right),  \nonumber \quad
\sigma_2=\left(\begin{array}{ccccc}
0 & -i\\
i &  0
\end{array} \right), \nonumber \quad
\sigma_3=\left(\begin{array}{ccccc}
1 & 0\\
0 & -1
\end{array} \right). \nonumber
$$
}
$$
V_2 = - \frac{i \l }{ 2} \sigma_3 +Y_0 =
\(\begin{array}{ccc}
 - \frac{i\l}{ 2} &  0\\
 0 & \frac{i\l}{ 2} \end{array}\)  +
\(\begin{array}{ccccc}
  0& \overline{\psi} \\
 \psi & 0 \end{array}\)
$$
and
$$
V_3 = \frac{\l^2}{ 2}i \sigma_3 -\l Y_0 + |\psi|^2 i\sigma_3 -i \sigma_3 Y_0'.
$$
We often omit the lower index and write $V=V_2$. 

The  NLS Hamiltonian $\HH=\HH_3$ is one in the infinite series of commuting  integrals of motion
\bey
\HH_1&=&\frac{1}{ 2} \int\limits_{\RB^1}  |\psi|^2 dx,\\
\HH_2&=&\frac{1}{ 2i} \int\limits_{\RB^1} \psib {\psi}'  dx,\\
\HH_3&=&\frac{1}{ 2} \int\limits_{\RB^1}  |\psi'|^2 +|\psi|^4 \, dx,\quad etc.\\
\eey
Hamiltonians produce an infinite hierarchy of  flows $e^{tX_m},\; m=1,2,\ldots$. 
Each flow $e^{tX_m}$ of the hierarchy can be written as the zero curvature condition 
with the suitable operator $\partial_t- V_m$. 

\subsection{Jost solutions}  
In this section we introduce the classical Jost solutions. These asymptotically normalized solutions play the central role in  analysis of the scattering spectral problem. Most of this material is standard but we present it here to fix notations.  We omit the time variable $t$ in all formulas of this section.

The   auxiliary linear problem     $(\partial_x  -V )\f=0, $  
can be written as an  eigenvalue problem for the  symmetric Dirac operator \ref{dirac}.  
The    transition  matrix $M(x,y,\l),\; x \geq y$;  satisfies  the differential equation 
$$
M'(x,y,\l)=V(x,\l) M(x,y,\l),
$$
and the boundary condition $ M(y,y,\l)=I$. The  transition  matrix is given by 
$$
M(x,y,\l)=\exp\int_{y}^{x} V(\xi,\l) d\xi.
$$
The   matrix $M(x,y,\l)$ is unimodular  because $V$ is traceless.  

Also, we introduce the reduced transition  matrix $T(x,y,\l),\, x\geq y;$ by the formula
\bay\label{red}
T(x,y,\l)=E^{-1}\(\frac{\l x}{ 2}\) M(x,y,\l) E^{-1}\(-\frac{\l y}{ 2}\),
\ey
where $E(\frac{\l x}{ 2})= \exp{(-\frac{i\lambda x}{ 2}  \sigma_3)}$ is a solution of
the linear problem with  $\psi\equiv 0$.  The matrix $T(x,y,\l)$ solves the equation
$$
T'(x,y,\l) = Y_0(x) E(\l x) T(x,y,\l),
$$
and satisfies the boundary condition $T(y,y,\lambda)= I$. 
Now the spectral parameter enters multiplicatively into the RHS of the differential equation.
The solution is given by the formula
\bay\label{exp}
T(x,y,\lambda)= \exp \, \int_{y}^x Y_0(\xi) E(\l\xi) d\xi. 
\ey
 The symmetry of the matrix
$Y_0:\, \sigma_1 Y_0(x)\sigma_1= \overline{Y_0(x)}$
is inherited by the  reduced transition matrix 
:$$\; \sigma_1T(x,y,\overline{\lambda})\sigma_1=
\overline{T(x,y,\lambda)}.$$
For  real  $\l$  formula \ref{exp} and  rapid decay of the potential imply an existence of the limit
$$
T(\lambda)= \lim T(x,y,\lambda)= \(\begin{array}{ccccc}
a(\lambda) & \overline{b}(\lambda)\\
b(\lambda) & \overline{a}(\lambda) \end{array} \), \quad
{\rm when} \quad y \rightarrow -\infty \quad\text{and}\quad x
\rightarrow +\infty;
$$
and  $|a(\lambda)|^2- |b(\lambda)|^2=1.$ Note, that $b(\l) \in S(\RB^1)$ since $\psi\in  S(\RB^1)$.

We introduce {\it Jost solutions}  $J_{\pm}(x,\lambda)$ as a matrix solutions of the differential equation
$$
J_{\pm}'(x,\lambda)=V(x,\lambda)J_{\pm}(x,\lambda),
$$
that also have prescribed   asymtotics at the spatial infinity
$$
J_{\pm}(x,\l)=
E\(\frac{\lambda x}{ 2}\) + o(1),\quad {\rm when}\quad x\rightarrow \pm \infty.
$$
An existence and analytic properties of the Jost solutions follow from the integral
representations
$$
J_+(x,\lambda)=E\(\frac{x\lambda}{ 2}\)+ \int_x^{+\infty} \Gamma_+(x,\xi)
E\(\frac{\lambda \xi}{ 2}\) d\xi,
$$
$$
J_-(x,\lambda)=E\(\frac{x\lambda}{ 2}\)+ \int_{-\infty}^x \Gamma_-(x,\xi)
E\(\frac{\lambda \xi}{ 2}\) d\xi.
$$
The kernels $\Gamma_{\pm}$ are unique and infinitely smooth in both variables.
Introducing the   notation $J_\pm=\[\jo^{(1)}_\pm,\, \jo^{(2)}_\pm\]$ we see from the
integral representations that
$\jo^{(1)}_-(x,\l),$ $  \jo^{(2)}_+(x,\l)$ are analytic in $\l$ in the upper half-plane
and continuous up to the boundary. Also, the columns
$\jo^{(2)}_-(x,\l),\;\; \jo^{(1)}_+(x,\l)$
are analytic  in the lower half--plane and continuous up to the boundary.

Now we describe analytic properties of the coefficient $a(\l)$  of the matrix $T(\l)$. 
The monodromy matrix $M(x,y,\l)$ can be written in the form
$$
M(x,y, \l)=J_+(x)J^{-1}_{+}(y)= J_{-}(x)J^{-1}_{-}(y).
$$
Therefore, 
$$
J_{+}^{-1}(x)\, M(x,y,\l)\, J_{-}(y)= J^{-1}_{+}(y)J_{-}(y)=J_{+}^{-1}(x)J_{-}(x).
$$
The variables $x$ and $y$ separate and  the above expression does not depend on $x$ or $y$ at all. 
By passing to the limit  with
$ x\rightarrow +\infty,\, y \rightarrow -\infty$ we have 
$$
T(\l)=J^{-1}_{+}(y)J_{-}(y)=J_{+}^{-1}(x)J_{-}(x).
$$
Therefore  $a(\l)= {\jo_-^{(1)}}^T(\l)J \jo_+^{(2)}(\l).$
The properties of Jost solutions imply that 
\begin{itemize}
\item  $a(\lambda)$ is analytic in the upper half-plane and continuous  up to the boundary;

\item   $a(\lambda)$ is  root-free;
\item $|a(\lambda)| \geq 1$ 
and $|a(\lambda)|^2-1 \in S(\RB^1)$ for $\l$ real, $a(\lambda) = 1 +o(1)$ as $ |\lambda| \longrightarrow \infty$. 
\end{itemize}

Let $p_\infty(\lambda)$ be  such that $a(\lambda)= \exp(-i2 p_{\infty}(\lambda))$
for $\lambda$ in the upper half-plane.  From the properties of $a(\lambda)$
it follow that 
\begin{itemize}
\item  $ \pin $  is analytic in the upper half-plane and
continuous  up to the boundary;  
\item  $\Im \pin \geq 0$ for $ \Im \lambda \geq  0$; 
\item  $\pin=o(1)$ for $|\lambda| \rightarrow \infty$;
for real $\lambda$,  the density of the measure $d \mi= \Im \pin \, d \lambda$ belongs to  
$S(\RB^1)$.  
\end{itemize}
The function $\pin$ can be written in the form
$$
\pin=\frac{1}{ \pi}\int \frac{d \mu_{\infty}(t)}{ t -\lambda}.
$$
Expanding the denominator in inverse powers of $\lambda$, we obtain:
\bay\label{V}
\pin= -\sum^{\infty}_{k=0}\frac{1}{ \lambda^{k+1}} \frac{1}{ \pi}\int_{-\infty}^{+
\infty} t^k d \mu_{\infty}(t)= -\frac{ \HH_1}{ \lambda}-  \frac{ \HH_2}{ \lambda^{2}}-
\frac{ \HH_3}{  \lambda^3}- \dots .
\ey
where $\HH_1,\HH_2$ and $\HH_3$ are the integrals introduced above.
The expansion has an asymptotic character for $\lambda: \; \delta \leq
\arg \lambda \leq \pi -  \delta,\;\delta > 0$.

To  describe asymptotic behavior in $x$  of the Jost solutions  $\jo_+^{(2)}(x,\l)$ and $\jo_-^{(1)}(x,\l)$ we assume that $\l$ is real and fixed. Then, we have the scattering rule
$$
\begin{array}{ccccccccccccccccc}
&\phantom{kkkkkkkkk} &x \rightarrow - \infty  &  &x\rightarrow  +  \infty \\
&\jo_+^{(2)}   &a(\l) \f_{\rightarrow}(x,\l)-\bar{b}(\l)\f_{\leftarrow}(x,\l)&   &
\quad \f_{\rightarrow}(x,\l)\\
&\jo_-^{(1)}   &\f_{\leftarrow}(x,\l) &\phantom{h} &
a(\l)\f_{\leftarrow}(x,\l)+
b(\l) \f_{\rightarrow}(x,\l),&
\end{array}
$$
where
$$
\f_{\leftarrow}(x,\l)=\left[ \begin{array}{ccccccccccccccccc}
                                           e^{- i\lt x}\\
                              0\end{array} \right], \quad
\f_{\rightarrow}(x,\l)=\left[ \begin{array}{ccccccccccccccccc}  0\\
                    e^{i\lt x}\end{array} \right]
$$ are  solutions of the free equation.
Similar for the Jost solutions $\jo_-^{(2)}(x,\l)$ and $\jo_+^{(1)}(x,\l)$ analytic in the lower half plane we assume that $\l$ is real and fixed. 
Then, we have the scattering rule
$$
\begin{array}{ccccccccccccccccc}
&\phantom{kkkkkkkkk} &x \rightarrow - \infty  &  &x\rightarrow  +  \infty \\
& \jo_+^{(1)} &\bar{a}(\l) \f_{\leftarrow}(x,\l)-{b}(\l)\f_{\rightarrow}(x,\l)&   &
\quad \f_{\leftarrow}(x,\l)\\
&\jo_-^{(2)}    &\f_{\rightarrow}(x,\l) &\phantom{h} &
\bar{a}(\l)\f_{\rightarrow}(x,\l)+
\bar{b}(\l) \f_{\leftarrow}(x,\l).&
\end{array}
$$

\begin{rem}\label{ffgr} 
If  $\f(x,\l)$ is a solution of the auxiliary problem 
$
(\partial_x  -V(x,\l) )\f=0
$
corresponding to $\l$, then $\hat \f=\sigma_1 \bar{\f}$ is a solution of $(\partial_x  -V(x,\overline{\l}) )\hat\f=0$ corresponding to 
$\overline{\l}$. For example, 
$\hat \fl(\l)= \fr({\bar \l})$. 
\end{rem}

We also introduce   the matrix BA function  
$$
H_+(\l)=\[ \jo_-^{(1)}(\l),\jo_+^{(2)}(\l)\] \quad {\rm and}\quad\quad H_-(\l)=\[ \jo_+^{(1)}(\l),\jo_-^{(2)} (\l) \]
$$ analytic in the upper/lower hulf-plane respectively. They are connected by the gluing condition
\bay\label{RH}
H_-(x, \l)= H_+(x,\l)  S(\l),
\ey
where  $\l \in \RB^1$ and the scattering matrix $S(\l)$
$$
S(\l) = \frac{1}{ a} \[\begin{array}{ccccc} \;\; {1} & {\overline{b}}\\
                       -{b}& {1}
                          \end{array}\].
$$
The gluing condition easily follows from the scattering rule.

Any Jost solution satisfies $\[J\partial_x- J V\]\jo=0$.
One can easily prove that $\jo^T$ satisfies
$\jo^T\[J\partial_x- J V\]=0.$ 
The matrices  $H_+^+$ and $H_-^+$ are  defined as
$$
H_+^+(\l)=\sigma_1 H_+^T(\l)= \[\begin{array}{ccccccc}   \jo_+^{(2)\,T}\\
                                     \jo_-^{(1)\,T}
              \end{array}\],  \qquad\qquad 
H_-^+(\l)=\sigma_1 H_-^T(\l)= \[\begin{array}{ccccccc}   \jo_-^{(2)\,T}\\
                                     \jo_+^{(1)\,T}
\end{array}\]. 
$$
Extending $a$ into the lower hulf-plane by the formula  $a^*(\l)= \overline{a(\bar \l)}$ we define
$$
H_+^*(\l)= -\frac{\sigma_3}{ a(\l)} H_+^+(\l),\qquad \qquad \qquad\qquad \text{for}\qquad \Im \l >0;
$$
and
$$
\qquad  H_-^*(\l)=  -\frac{\sigma_3}{ a^*(\l)} H_-^+(\l),\qquad \qquad \qquad \text{for}\qquad \Im \l < 0.
$$
Then the   dual  gluing condition holds
\bay\label{ARH}
H_-^*(x,\l) =S^{-1}(\l) H_+^*(x,\l),\qquad\qquad\qquad{\rm where} \qquad \l\in \RB^1,
\ey
and
$$ S^{-1}(\l)= \frac{1}{ a^*} 
\[\begin{array}{ccccc} \;\; {1} & -{\overline{b}}\\
                        \;\;{b}& {1}
    \end{array}\].
$$

Next  lemma provides explicitly a few terms of the  asymptotic  expansion of  Jost solutions.
\begin{lem}\label{assu} (i) For  fixed $x$ the following formulas hold
$$
\jo_+^{(2)}(x,\l)=
e^{+i \lt x} \sum\limits_{s=0}^{\infty} \[\begin{array}{cccc}
 g_s \\
 k_s
\end{array} \] \l^{-s}, 
$$
 and
$$
\jo_-^{(1)}(x,\l)
= e^{-i \lt x} \sum\limits_{s=0}^{\infty} \[\begin{array}{cccc}
  h_s \\
  f_s
\end{array} \] \l^{-s} , 
$$
 The expansion has an asymptotic character for
 $\l:\; \d \leq \arg \l \leq \pi -\d,\;\d >0$.

 (ii) For  fixed $x$ the following formulas hold
$$
\jo_+^{(1)}(x,\l)=
e^{-i \lt x} \sum\limits_{s=0}^{\infty} \[\begin{array}{cccc}
 \bar{k}_s \\
 \bar{g}_s
\end{array} \] \l^{-s},
$$
 and
$$
\jo_-^{(2)}(x,\l)= e^{+i \lt x} \sum\limits_{s=0}^{\infty} \[\begin{array}{cccc}
 \bar f_s \\
 \bar h_s
\end{array} \] \l^{-s}. 
$$
The expansion has an asymptotic character for
 $ \l:\;-  \d \geq \arg \l \geq -\pi + \d,\; \d >0$.

(iii) The coefficients $g$'s and $k$'s are given by the formulas
$$
g_0= 0,\quad \quad \quad \quad
g_1= -i \psib,
$$
and
$$
 k_0= 1, \quad \quad \quad \quad
 k_1 = i \int^{+\infty}_{x} |\psi(x')|^2 dx'.
$$

The coefficients $h$'s and  $ f$'s are given by the formulas
$$
f_0= 0,\quad \quad \quad \quad
f_1= i\psi,
$$
and
$$
 h_0= 1 , \quad \quad \quad \quad
 h_1 =  i \int_{-\infty}^{x} |\psi(x')|^2 dx'.
$$
\end{lem}

\begin{lem} \label{ppr}
The scattering map 
$$
\psi(x),\;x\in \RB^1 \quad \longrightarrow\quad b(\l),\; \l \in \RB^1
$$
is injective. 
\end{lem}
\noi
{\it Proof.} \cite{DT}. Assume that there are two different potentials with the same function $b(\l)$. Then the difference $\Delta \jo(x,Q), \; Q=(\l+i0,+);$ of the corresponding Jost solutions does  not vanish identically in the variable $ \l$ for some $x$. 
Using the scattering rule and Remark \ref{ffgr} we have 
$$
\hat {\jo}(x,Q)=-\frac{b}{ a}\jo(x,Q) + \frac{1}{ a} \jo(x,\epm Q).
$$
This identity produces
$$
\sigma_1 \Delta \bar{\jo}(x,Q)+\frac{b}{ a}\Delta \jo(x,Q) = \frac{1}{ a}\Delta \jo(x,\epm Q).
$$ 
Multiplying on $\Delta \jo^T(x,Q)\sigma_1$ from the left  
$$
|\Delta \jo(x,Q)|^2 +\frac{b}{ a}\Delta \jo^T(x,Q)\sigma_1\Delta \jo(x,Q) = \frac{1}{ a}\Delta \jo^T(x,Q)\sigma_1\Delta \jo(x,\epm Q).
$$
For arbitrary fixed $x$ the RHS is analytic in the upper half plane and decay there as $O(|\l(Q)|^{-2})$. 
Therefore by the Cauchy theorem 
$$
\int d\l\, |\Delta \jo(x,Q)|^2 +\int d\l\, \frac{b}{ a}\Delta \jo^T(x,Q)\sigma_1\Delta \jo(x,Q) = 0.
$$
Since 
$$
\frac {|b(\l)|}{|a(\l)|}  <1,
$$
the second term can not balance the first. The contradiction implies the result. 
\qed

We conclude our discussion of the Jost solutions with  the following lemma 
\begin{lem}\label{vjs} The variational derivatives of the Jost solution $\jo_+^{(2)}$
are given by the formulas\footnote{We abuse the notations denoting the components of the Jost solutions by upper indexes.}
\bey
\frac{\d \jo_+^{(2)}(x)}{\d \psi(y)} &=&\frac{\d \jo_+^{(2)}(x)}{\d \psib(y)}=0,\quad\qquad\qquad\qquad\qquad\qquad\qquad y < x; \\
\frac{\d \jo_+^{(2)}(x)}{\d \psi(y)} &=& -\frac{j_-^1j_+^1(y)}{a}\jo_+^{(2)}(x) + \frac{j_+^1j_+^1(y)}{a}\jo_-^{(1)}(x), \qquad\qquad y > x; \\
\frac{\d \jo_+^{(2)}(x)}{\d \psib(y)} &=& + \frac{j_+^2j_-^2(y)}{a}\jo_+^{(2)}(x) - \frac{j_+^2j_+^2(y)}{a}\jo_-^{(1)}(x), \qquad\qquad y > x.
\eey
The variational derivatives of the Jost solution $\jo_-^{(1)}$
are given by the formulas
\bey
\frac{\d \jo_-^{(1)}(x)}{\d \psi(y)} &=&\frac{\d \jo_-^{(1)}(x)}{\d \psib(y)}=0,\quad\qquad\qquad\qquad\qquad\qquad\qquad  x <y ; \\
\frac{\d \jo_-^{(1)}(x)}{\d \psi(y)} &=& -\frac{j_-^1j_+^1(y)}{a}\jo_-^{(1)}(x) + \frac{j_-^1j_-^1(y)}{a}\jo_+^{(2)}(x), \qquad\qquad y < x; \\
\frac{\d \jo_-^{(1)}(x)}{\d \psib(y)} &=& +\frac{j_+^2j_-^2(y)}{a}\jo_-^{(1)}(x) - \frac{j_-^2j_-^2(y)}{a}\jo_+^{(2)}(x), \qquad\qquad y < x.
\eey
\end{lem}
{\it Proof.} We give complete proof for the first set of formulas. The second can be proved  in the same way. 

 Let ${\jo^{(2)}_+}^\bullet$ be a variation of $\jo_+^{(2)}$ in response to the variation of $\psi$.  
Then  
$$
{{\jo_{+}^{(2)}}^{\bullet}} '=V{\jo_+^{(2)}}^\bullet+  V^\bullet \jo_+^{(2)},
$$
where $V^\bullet$ is a variation of $V$. Then, it can be readily verified that  
$$
{\jo_+^{(2)}}^\bullet= -\int\limits_{x}^{+\infty} d\xi\, H(x)H^{-1}(\xi)V(\xi)\jo_+^{(2)}(\xi),
$$
where $H=H_+(x,\l)=\[\jo_-^{(1)},\jo_+^{(2)}\]$. From this one reads, 
$$
\frac{\d \jo_+^{(2)}(x)}{\d \psi(y)}=- H(x)H^{-1}(y)  \left(\begin{array}{ccccc}
0 & 0\\
1 & 0
\end{array} \right)  \jo_+^{(2)}(y),\qquad\qquad \qquad  y>x; 
$$
and 
$$
\frac{\d \jo_+^{(2)}(x)}{\d \psib(y)}=- H(x)H^{-1}(y)  \left(\begin{array}{ccccc}
0 & 1\\
0 & 0
\end{array} \right)  \jo_+^{(2)}(y),\qquad\qquad \qquad  y>x. 
$$
Using the scattering rule one computes $\det H(x,\l)=a(\l)$ and inverts the matrix $H(y)$. The straightforward computation produces the result. 
\qed

\section{The spectral cover and meromorphic functions on it. }
\subsection{The spectral cover. } 

The spectral cover   and the  Weyl functions  for general potentials were  considered  in \cite{V3}. 
In this section we introduce two other functions $\Pi$ and $\U$ which important in the discussion of the scattering case.

Evidently the Jost solutions are particular case of the general Weyl solutions  normalized not at some  finite point   
but asymptotically when  $x\longrightarrow \pm \infty$.  
Therefore,  the construction of the Weyl function on the spectral cover  can be carried for the scattering case without changes. 
We present this construction here because it will be  also needed for the construction of the functions $\Pi$ and $\U$. 

For each point of the complex plane of the spectral parameter $\l$  with nonzero imaginary part there are two Jost solutions. 
To make the Jost solution a single valued function of a point we introduce $\G=\G_{+} \cup \G_{-}$, a two sheeted covering of the 
complex plane (Figure 2). Each sheet $\G_{+}$ or $\G_{-}$ is a copy of the complex plane cut along the real line. 
Each point  of the cover $\G$ is a pair $Q=(\l,\pm)$ where $\l$ is a point of the complex plane and the sign $\pm$ specifies the sheet. 
We denote by $P_+$ or $P_-$ the  infinity corresponding to the   sheet $\G_{+}$ or $\G_{-}$. 
\begin{figure}[htb]
\includegraphics[width=0.60\textwidth]{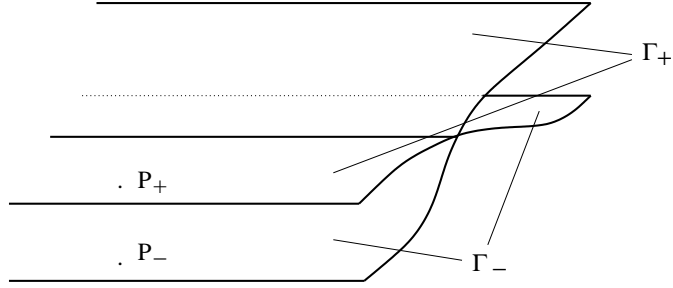}
\caption{ Two sheets of  the spectral cover.}
\end{figure}

\noi
Let us introduce two components of the spectral cover
$$
\G_R=\{Q\in \G_+,\; \Im \l(Q) >0\} \cup \{Q\in \G_-,\; \Im \l(Q) <0\}
$$
and 
$$
\G_L=\{Q\in \G_+,\; \Im \l(Q) <0\} \cup \{Q\in \G_-,\; \Im \l(Q) > 0\}.
$$
Evidently $\G=\G_R\cup \G_L$. 

On $\G$ we define an involution $\epm$ permuting sheets by the rule 
$$
\epm:\quad (\l,\pm)\longrightarrow (\l,\mp).
$$
Obviously, the involution permutes infinities $\ea: P_+\longrightarrow P_-$ but it  leaves $\G_R$ and $\G_L$ invariant. 
On $\G$ we also define an involution $\ea$ by the rule 
$$
\ea:\quad (\l,\pm)\longrightarrow (\lb,\mp).
$$
The involution permutes infinities $\ea: P_+\longrightarrow P_-$ and commutes with $\epm$. 

Now we lift Jost solutions from the $\l$--plane of the spectral parameter
to the spectral cover, where they become single valued function of a point. We define

\noi
for $Q\in \G_+$:
$$
\jo(x,Q)= \left\{  \begin{array}{ll}
                \jo^{(2)}_+(x,\l)& \qquad\qquad\mbox{ if $\Im \l(Q) >0$},\\
                \jo^{(2)}_-(x,\l)&\qquad\qquad \mbox{ if $\Im \l(Q) <0$};
                    \end{array}
                    \right.
$$
for $Q\in \G_-$:
$$
\jo(x,Q)= \left\{  \begin{array}{ll}
                \jo^{(1)}_-(x,\l)&\qquad\qquad \mbox{ if $\Im \l(Q) >0$},\\
                \jo^{(1)}_+(x,\l)&\qquad\qquad \mbox{ if $\Im \l(Q) <0$}.
                    \end{array}
                    \right.
$$

Let  $\f^T$ denote the transposition of the vector $\f$ and let $\f^*$ denote the adjoint of the vector $\f$. Let $\LLL^2(a,b)$ be a space of vector functions with the property
$$
\int\limits_{a}^{b} \f^*(x,\l)\f(x,\l) \,dx < \infty.
$$ 
As it follows from the construction\footnote{The parameter $y$ is an arbitrary real number.} $\jo(x,Q)\in \LLL^2[y,+\infty)$ when $Q\in \G_R$ and $\jo(x,Q)\in \LLL^2(-\infty,y]$ when $Q\in \G_L$.  

For  general potentials considered in \cite{V2} a rule connecting Weyl solutions on different banks of the cut is not specified.  For the periodic potentials considered in \cite{V3} the suitably normalized Weyl (Floquet) solutions can be analytically extended across the real line.   
On the contrary, in the scattering case this can not be done, 
but the  gluing condition \ref{RH} connects different branches of the Jost function $\jo(x,Q)$. 

The next Lemma   shows that  the properties of $\jo(x,Q)$ resemble properties  of the Floquet solutions   for periodic potentials 
(see Lemma 2.5, \cite{V3}).  
\begin{lem}\label{basc} The following identity holds 
\bay\label{conju}
\jo(x,\ea Q)=\sigma_1 \overline{\jo(x,Q)}.
\ey

In the vicinity of  infinities  the  function $\jo(x,Q)$ has the  asymptotic, 
\bay\label{J}
\jo(x,Q) =  e^{\pm i\frac{\l}{ 2}x} \[\jo_0/\hat \jo_0  + o\(1\)\], \qquad\qquad\qquad Q\in(P_{\pm}),\quad \l=\l(Q), 
\ey
and 
$$
\jo_0= \[\begin{array}{ccc} 0\\1\end{array}\],\qquad\qquad\qquad 
\hat \jo_0= \[\begin{array}{ccc} 1\\0\end{array}\].
$$
\end{lem}

\noi
{\it Proof. } The asymptotics \ref{J} follows from Lemma \ref{assu}. 

To prove \ref{conju} 
consider the solution $\sigma_1 \overline{\jo(x, Q)}$. Due to Remark \ref{ffgr} it must be a linear combination of two Jost solutions at the points above 
$\l(\ea Q)$.  
Using \ref{J} and comparing the asymptotics for $x \rightarrow \pm \infty$ we obtain the result.  \qed

  As in \cite{V2} we define the Weyl function by the formula 
$$
\X(x,Q)=\frac{\jo_2(x,Q)}{\jo_1(x,Q)},\qquad\qquad\qquad\qquad Q\in \G. 
$$
This is the same Weyl function that we introduced  using  general Weyl solution. It follows from \ref{conju} that 
$$
\X(x,\ea Q)=\frac{1}{\overline{\X(x, Q)}}.
$$
Thus we constructed  the map 
$$
\MM\quad  \longrightarrow \quad(\G,\X)
$$
which is called the direct spectral transform. 

\subsection {The functions $\Pi$ and $\U$.}  In this section we construct  meromorphic functions $\Pi$ and $\U$, following ideas of \cite{V3, V4}. 

First let us introduce the dual Jost solution $\jo^+(x,Q)$  by the formula
$$
\jo^+(x,Q)=\jo^T(x,\epm Q). 
$$
The gluing condition \ref{ARH}  connect various branches of the dual Jost solution. 

For any Jost solution or dual Jost solution   we define 
$$
\js(x,Q)=\oneb^T \jo(x,Q),\qquad\qquad\qquad \js^+(x,Q)=\jo^+(x,Q)\oneb, 
$$  
where $\oneb^T=(1,1)$. Due to the fact that   the Dirac spectral problem is symmetric the function $ \js(x,Q)$ or $\js^+(x,Q)$  
never vanishes for any $x$ and $Q$ with  $\Im \l(Q) \neq 0$.

Let us introduce the Wronskian function $\W(Q)$ by the formula 
$$
\W(Q)=\jo^+(x,Q) J \jo(x,Q).  
$$It can be proved easily $\W(\epm Q)=-\W(Q)$ and  $\W(\ea Q)=-\overline{\W(Q)}$. From the asymptotic formulas for $Q \in \G_+$ with $\Im \l(Q) > 0$ 
we have $\W(Q)=a(\l)$. It can be computed  on  other parts of the cover using involutions. 

Let us introduce the  function $\PP(x,Q)$ by
$$
\PP(x,Q)= \js_+(x,Q) \js_-(x,\epm Q). 
$$
Apparently, $\PP(x,\epm Q)= \PP(x,Q)$ and it  depends on $\l=\l(Q)$ only; $\PP(x,\ea Q)=\overline{\PP(x,Q)}.$

We introduce the  function $\Pi(x,Q)$   by the formula:
$$
\Pi(x,Q)=\frac{\PP(x,Q)}{\W(Q)}.
$$
We have $\Pi(x,\epm Q)= -\Pi(x,Q)$ and  $\Pi(x,\ea Q)= -\overline{\Pi(x,Q)}$. Since dependence on the sheet is very simple we introduce 
the function $\Pi(x,\l)$ which coincides with  $\Pi(x,Q)$ on $\G_+$. Apparently, 
$$
\Pi(x,\l)=\frac{\js_+(x,Q) \js_-(x,\epm Q)}{ a(\l)},\qquad\qquad \Im \l=\l(Q) >0;
$$
and
$$
\Pi(x,\l)=\frac{\js_+(x,Q) \js_-(x,\epm Q)}{ a^*(\l)},\qquad\qquad \Im \l=\l(Q) < 0.
$$
The  function $\Pi(x,\l)$ is defined on the complex plane cut along the real line.
Its' properties  follow from Lemmas \ref{assu},   and \ref{ffgr}. 

\begin{itemize} 
\item The holomorphic function $\Pi(x,\l)$ never vanishes  on the complex plane cut along the real line. It satisfies the identity
\beq\label{conpi}
\Pi(x,\lb)=\overline{\Pi(x,\l)}.
\eeq
\item It has the following asymptotics for $\l$ such that $\delta \leq \arg \l \leq \pi -\delta,\; \delta >0$, 
$$
\Pi(x,\l)=  1+ \frac{i\psi(x)-i\psib(x)}{\l}+\hdots,
$$ and in the angle $-\delta > \arg \l > -\pi +\delta,\; \delta >0,$ 
$$
\Pi(x,\l)=1+ \frac{i\psi(x)-i\psib(x)}{\l}+\hdots.
$$
\end{itemize}
We deal with $\Pi(x,Q)$ or $\Pi(x,\l)$ depending on the situation. 

We introduce the function $\U(x,Q)$ by the formula 
$$
\U(x,Q)=\frac{\js(x,Q)}{ \js(x,\epm Q)},\qquad\qquad\qquad Q\in \G. 
$$
Evidently,
$$
\U(x,\epm Q)=\frac{1}{\U(x, Q)}
$$
and 
$$
\U(x,\ea Q)=\overline{\U(x, Q)}.
$$
The  function $\U(x,\l)$ is defined on the complex plane cut along the real line and coincides with $\U(x,Q)$ on $\G_+$. It has the following properties 
\begin{itemize} 
\item The holomorphic function $\U(x,\l)$ never vanishes  on the complex plane for $\Im \l \neq 0$. It satisfies the identity
\beq\label{conup}
\U(x,\lb)=\frac{1}{\overline{\U(x,\l)}}.
\eeq
\item It has the following asymptotics\footnote{The anti-derivative $D^{-1}$ is defined by the formula $D^{-1}f(x)=\frac{1}{2} \[
\int\limits_{-\infty}^{x}dy f(y)- \int\limits^{+\infty}_{x}dy f(y)\]$.}   for $\l$ such that $\delta \leq \arg \l \leq \pi -\delta,\; \delta >0$,
$$
\U(x,\l)=  e^{i\l x}\[1+ \frac{-i\psi(x)-i\psib(x)-2iD^{-1} |\psi|^2(x)}{\l}+\hdots\], 
$$ and in the angle $-\delta > \arg \l > -\pi +\delta,\; \delta >0,$ 
$$
\U(x,\l)=  e^{i\l x}\[1+ \frac{-i\psi(x)-i\psib(x)-2iD^{-1} |\psi|^2(x)}{\l}+\hdots\]. 
$$
\end{itemize}


Since $\Pi(x,\l)$ and $\U(x,\l)$ are holomophic in the open upper and lower half planes and do not have zeros there,  we can define 
  $\Omega$ and $\Xi$ to be  the logarithm   of these functions: 
\begin{eqnarray*}
\U&=& e^{\Omega}= e^{\c +i\ct},\\
\Pi&=& e^{\Xi}=e^{\xit +i \xi}.
\end{eqnarray*}
The functions $\Pi(x,\l)$ and $\U(x,\l)$ take different values when $\l$ approaches the 
real axis from above and below.  
The identities \ref{conpi} and \ref{conup}  immediately imply
$$
\c(x,\l-i0)=-\c(x,\l+i0),\qquad \qquad \ct(x,\l-i0)=\ct(x,\l+i0);
$$
and
$$
\xi(x,\l-i0)=-\xi(x,\l+i0),\qquad \qquad \xit(x,\l-i0)=\xit(x,\l+i0).
$$
\subsection{The function $\Pi$ and the resolvent. } 
In this subsection we alternatively define the function $\Pi$ using the resolvent of the 
symmetric operator $\Dg$. Indeed,  $\Dg- \lt$ can be inverted for $\l,\; \Im \l \neq 0$, {\it i.e.}   
if $\(\Dg-\lt\)\f=\e$, then 
$
\f=\Rg(\l)\e,
$ where $\Rg=\(\Dg-\lt\)^{-1}$ is the resolvent.  

The resolvent  is an integrable operator given by the formula 
$$
\f(x)=\Rg(\l)\e=\int\limits_{-\infty}^{+\infty} R(x,y,\l) \e(y) dy.
$$
For $\l$ with $\Im \l > 0$ and $Q\in \G_+$ above  this $\l$ the kernel $R(x,y,\l)$ is given by the formulas 
$$
R(x,y,\l)= \frac{\jo(x,Q) \jo^T(y,\epm Q)i\sigma_1}{ a(\l)},\qquad\qquad
\qquad\mbox{for}\qquad y\leq x;
$$
$$
R(x,y,\l)=  \frac{\jo(x,\epm Q) \jo^T(y,Q)i\sigma_1}{ a(\l)},\qquad\qquad
\qquad\mbox{for}\qquad y\geq x.
$$

\noi
For $\l$ with $\Im \l < 0$ and $Q\in \G_+$ above  this $\l$  the kernel $R(x,y,\l)$ is given by the formulas 
$$
R(x,y,\l)= -\frac{\jo(x,\epm Q) \jo^T(y,Q)i\sigma_1}{ a^*(\l)},\qquad\qquad
\qquad\mbox{for}\qquad y\leq x;
$$
$$
R(x,y,\l)=  -\frac{\jo(x,Q) \jo^T(y,\epm Q)i\sigma_1}{ a^*(\l)},\qquad\qquad
\qquad\mbox{for}\qquad y\geq x.
$$

Let us introduce the vector-function $\db_x$ as 
$\db^T_x(y)=(\delta_1(x-y),\delta_2(x-y))$.   
The formula for the resolvent kernel   implies\footnote{The sign $<\quad, \quad >$ signifies the inner product in $L^2(\RB^1)$.}   
$$
 i \Pi(x,\l)=<\db_x, \Rg(\l)\db_x>,\qquad\qquad\qquad\Im \l > 0,
$$
and 
$$
 -i \Pi(x,\l)=<\db_x, \Rg(\l)\db_x>,\qquad\qquad\qquad\Im \l < 0.
$$
This formula is quite suggestive.  It was shown in \cite{V0}  the Poisson bracket for the function similar to $\Pi$ can be computed in terms of the function itself. Here the situation is more complicated as it will be  demonstrated in Theorem \ref{pbpi}. 
\begin{rem} These formulas imply  that 
$$
\pm \frac{\pi}{2} +\xi(x,\l)= \Im \log <\db_x, \Rg(\l)\db_x>,\qquad\qquad\qquad \Im \l >0/\Im \l <0.
$$
Similar formula was used by Krein, \cite{Kr}, to define by means of  the resolvent the, so--called, spectral shift function. See also \cite{SG} for 
the trace formulas using the  Krein's  spectral shift.  
\end{rem}

\newpage
\section{ The  Scattering divisor and the Abel map.} 
\subsection{ Useful identities.} 
If it is not 
mentioned  otherwise  we always assume that $\l$ lies on the upper bank, 
{\it i.e.} $\l=\l+i0$. We derive  two identities that will be used  in  calculations.
\begin{thm}
For $\l=\l+i 0$ and any $x \in \RB$ the following identities  hold:
\bay
|a|e^{-i\xi}&=& -b e^{i\ct}+ e^{-\c},\label{first}\\ \label{second}
|a|e^{-i\xi}&=&e^{\c}+\bb e^{-i\ct}, 
\ey 
where  $a=a(\l),\; \xi=\xi(x,\l)$, etc.
\end{thm} 

\noi
{\it Proof.}   Let $Q=(\l+i0,+)$. Using the scattering rule and Remark \ref{ffgr} we have 
$$
\hat {\jo}(x,Q)=-\frac{b}{ a}\jo(x,Q) + \frac{1}{ a} \jo(x,\epm Q).
$$
Therefore,
$$
\bar {\js}(x,Q)=-\frac{b}{ a}\js(x,Q) + \frac{1}{ a} \js(x,\epm Q).
$$
Multiplying  on $\js(x,Q)$ we obtain 
\bay\label{gprs}
| {\js}(x,Q)|^2=-\frac{b}{ a}\js^2(x,Q) + \frac{\js(x,Q)\js(x,\epm Q)}{ a} .
\ey
The formula 
$$\frac{\js^2(x,Q)}{ a(\l)} = \Pi(x,Q) \U(x,Q)=
e^{\xit + i\xi +\c + i\ct}
$$
implies, in particular,  $|\js(x,Q)|^2=|a|e^{\xit +\c}$. From these  we obtain \ref{first}.  

The proof of \ref{second} is similar. Using the scattering rule and Remark \ref{ffgr} we have 
$$
\hat {\jo}(x,\epm Q)=\frac{\bar{b}}{ a}\jo(x,\epm Q) + \frac{1}{ a} \jo(x, Q). 
$$
Therefore,
$$
\bar {\js}(x,\epm Q)=\frac{\bar{b}}{ a}\js(x,\epm Q) + \frac{1}{ a} \js(x, Q). 
$$
Multiplying  on $\js(x,\epm Q)$ 
$$
| {\js}(x,\epm Q)|^2=\frac{\bar{b}}{ a}\js^2(x,\epm Q) + \frac{\js(x, Q)\js(x,\epm Q)}{ a} , 
$$
and using the formula 
$$\frac{\js^2(x,\epm Q)}{ a} = \frac{\Pi(x,Q)}{\U(x,Q)}=
e^{\xit + i\xi -\c - i\ct}$$
which implies, in particular, $|\js(x,\epm Q)|^2=|a|e^{\xit-\c}$ we obtain \ref{second}. 
\qed

The formulas \ref{first}--\ref{second} contain three functions $\c,\; \xi$ and $\ct$. Eliminating any of this functions  leads to three important identities.
 
The first relation  for   two functions $\c$ and $\xi$  
was derived first by Ercolani and  McKean, \cite{EM}:
\bay\label{mce}
|a|\cos \xi   =\cosh \c .
\ey
It is obtained by eliminating $\ct$ from the identities \ref{first}--\ref{second}. 
Since $\cosh$ and $\cos$ are even functions  formula \ref{mce} holds for the functions $\xi$ and $\c$ 
from both sides of the cut.

The second  important identity can be obtained by eliminating $\c$ from the 
formulae \ref{first}--\ref{second}. Taking  sum 
$$
|a|e^{-i\xi}=\cosh \c -i|b| \sin (\ct + \ph b).
$$
Computing the  imaginary  part of this formula we have 
$$
|a|\sin \xi=|b| \sin  (\ct + \ph b).
$$
After simple transformations we arrive at the formula
\beq\label{vam}
\ph \,b=-\ct +\sin^{-1}\left|\frac{a}{ b}\right| \sin \xi.
\eeq
This is a continuum version of the Abel map, first obtained by Venakides, \cite{Ven}. More ''geometric'' form of this formula will be given in \ref{sdiv}.

The last  third identity can be obtained by eliminating $\xi$ from the 
formulas \ref{first}--\ref{second}. Subtracting one from another 
$$
e^{\c}-e^{-\c}+\bb e^{-i\ct}+b e^{i\ct}=0,
$$
and
$$
\sinh \c +|b|\cos (\ct + \ph b)=0.
$$
After simple transformations we arrive at the formula
$$
\ph \,b=-\ct +\cos^{-1}-\frac{1}{ |b|}\sinh \c.
$$
Evidently, with  the aid of \ref{mce} changing the last 
term, we  obtain the Venakides formula.

\subsection{The scattering curve and divisor.}\label{sdiv}

The scattering curve  $\Gin$  was introduced by McKean and Ercolani in \cite{EM} for the KdV equation. It is  a continuum analog of the real ovals accommodating points of the divisor (poles of the Floquet solutions) in the periodic case \cite{V4}. This construction is considered here for the 
NLS. 

We assume that the function $a(\l)$ is known. The function   $h(\l)=\cos^{-1} |a(\l)|^{-1}$ determines the branch points of the  scattering curve $\Gin$. The scattering curve itself  has a continuum of the real ovals as it is shown on Figure 3.
For each $\l$ the real oval is two sheeted covering of the segment $[-h(\l),\;h(\l)]$.  The identity  \ref{mce} implies that the function $\xi(x,\l)$ satisfies the inequality
$$
-h(\l)\leq \xi(x,\l)\leq h(\l).
$$

\begin{figure}
\centering
\includegraphics[width=0.60\textwidth]{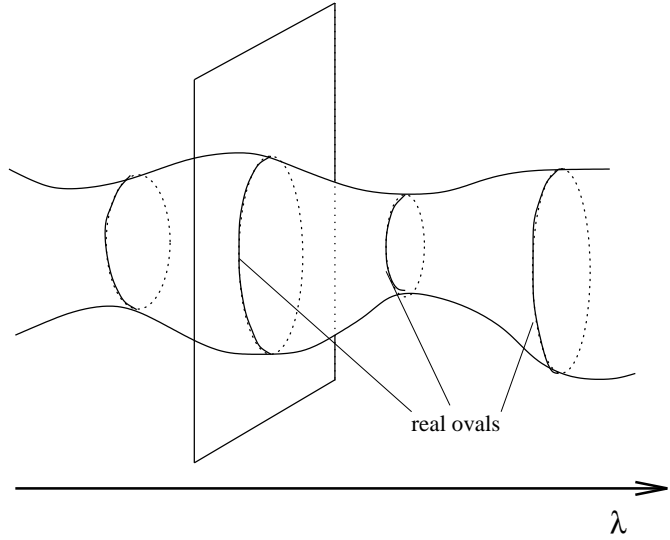}
\caption{The continuum of real ovals.}
\end{figure}

\noi
The set of pairs 
$$
\g(\l)=(\xi(x,\l), \; \sign \c(x,\l)),\qquad \l \in \RB^1;
$$
constitutes the, so--called,  scattering S-divisor. 
The coordinate $\xi(x,\l)$ 
determines the projection, while the $\sign \c(x,\l)$ determines the "+" or "-" arch of the oval as  on  Figure 4. 
The situation is quite similar to the periodic case, \cite{MV}. 


\begin{lem} 
Let us assume that the scattering curve (the function $|a(\l)|$) is fixed. The   S-divisor  determines the potential.
\end{lem}
\noi
{\it Proof.} Identity \ref{mce} determines the function $\c(\l),\; \l\in \RB^1$. Both harmonic functions $\c$ and $\xi$ can be recovered from 
 their values on the real line. Then the conjugate functions $\xit$ and $\ct$ also can be found and the additive constants are determined from the asymptotic for  the functions 
$\Pi(x,\l)$ and $\U(x,\l)$. Thus  $\js^2(x,Q)$ is known  from $\Pi$ and $\U$ on all parts of the spectral cover. The formula \ref{gprs} for $Q=(\l+i0,+)$
$$
| {\js}(x,Q)|^2=-\frac{b}{ a}\js^2(x,Q) + \Pi(x,\l), 
$$
provides the value of $b(\l)$ for all 
$\l\in \RB^1$.  Lemma \ref{ppr} implies the result. 

\qed

\begin{figure}
\centering
\includegraphics[width=0.60\textwidth]{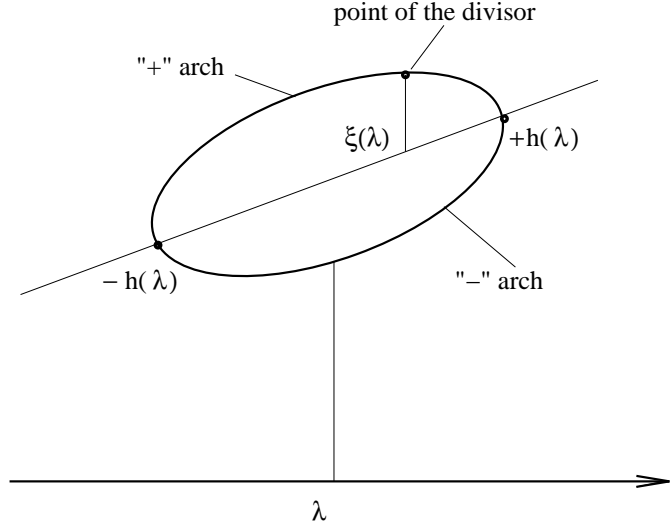}
\caption{The point of the S-divisor on the  real oval.}
\end{figure}

Now we  put formula \ref{vam} into more geometric form. We  fix some divisor $\g_0=(\xi_0,\sign \c_0)$ on the curve $\Gin$.  The formula \ref{mce} implies 
for each $x$ and $\l$ 
$$
\c-\c_0=\int_{\xi_0}^{\xi} \frac{d \cos \xi}{\sqrt{\cos^2\xi-\frac{1}{|a|^2}}}.
$$
The sign of the radical is the same as the sign of $\c$. We replace the limits of integrations on $\g$ assuming this agreement
$$
\c-\c_0=\int_{\g_0}^{\g} \frac{d \cos \xi}{\sqrt{\cos^2\xi-\frac{1}{|a|^2}}}.
$$
Similar,
$$
\sin^{-1} \left|\frac{a}{b}\right| \sin \xi- \sin^{-1} \left|\frac{a}{b}\right| \sin \xi_0=\int_{\g_0}^{\g} \frac{d \sin \xi}{\sqrt{\cos^2\xi-\frac{1}{|a|^2}}}.
$$
Therefore, denoting the Hilbert transform by $H$, we have
$$
\ph b- \ph b_0= -H\[ \int_{\g_0}^{\g} \frac{d \cos \xi}{\sqrt{\cos^2\xi-\frac{1}{|a|^2}}} \] + \int_{\g_0}^{\g} \frac{d \sin \xi}{\sqrt{\cos^2\xi-\frac{1}{|a|^2}}}.
$$
This form of \ref{vam} resembles the   Abel sum in the Baker's  form for the periodic case. The summation over open gaps is replaced by the singular integral 
of the Hilbert transform. 
The only difference  is the second term. This term corresponds to the contribution into the Abel sum of the  open gap which  does not contains zero 
of the normalized differential, see \cite{MV}. 

\subsection{Action of the NLS flows on $S$--divisor.} The main results of this section are Theorems \ref{divfl} and \ref{divfll} which describe the action of the first two flows of the NLS hierarchy on the scattering divisor. 

\begin{thm} \label{divfl}
For any $x\in \RB^1$ the action of the first vector field $X_1=\{\bullet, \HH_1\}$ on the S-divisor is given by the formula
\beq
X_1\xi(x,\l)=\frac{1}{|a(\l)| e^{\xit(x,\l)}}\sinh \c(x,\l).
\eeq
\end{thm}

  Using  formula \ref{mce} one can put  the result in the form
$$
X_1\xi(x,\l)=\pm e^{-\xit(x,\l)}\sqrt{\cos^2 \xi(x,\l)-\frac{1}{|a(\l)|^2}}.
$$
The sign of the radical is the same as   $\sign \c(x,\l)$. 
The instant  the function $\xi(x,\l)$ reaches the end (branch point) of the interval  $[-h(\l),\;h(\l)]$ the signature of the radical changes and  the direction of motion  reverses. 

We precede the proof  with the following

\begin{lem}\label{hhg}
The action of the first vector field $X_1=\{\bullet, \HH_1\}$ on the Jost solutions is given by the formulas
\bey
X_1 j_+^1(x) &=&i j_+^1(x),\qquad\qquad\qquad\qquad X_1j_-^1(x)=0,\\
X_1 j_+^2(x) &=& 0,\qquad\qquad\qquad\qquad\qquad X_1j_-^2(x)=-i j_-^2(x).
\eey
\end{lem}
\noi
{\it Proof.} We will present  complete proof of the formulas for the action of $X_1$ on $\jo_-(x)$. 
$$
X_1\jo_-(x)=\{\jo_-(x),\HH_1\}=2i\int\limits_{-\infty}^{+\infty} \frac{\d \jo_-(x)}{\d \psib(y)}\frac{\d \HH_1}{\delta \psi(y)}- 
\frac{\d \jo_-(x)}{\d \psi(y)}\frac{\d \HH_1}{\delta \psib(y)}   dy.
$$
Substituting the gradients from Lemma \ref{vjs} after simple algebra we obtain 
$$
\therefore\;=\frac{i\jo_-(x)}{a(\l)} \int\limits^{x}_{-\infty} \( j_-^2j^2_+\psib(y) + j_-^1j_+^1\psi(y)\) dy - \frac{i\jo_+(x)}{a(\l)} \int\limits^{x}_{-\infty} \( j_-^2j^2_-\psib(y) + j_-^1j_-^1\psi(y)\) dy.
$$

Both expressions  under the integral  signs turn out to be a full derivatives 
\bay\label{ppi} 
j_-^2j^2_+\psib + j_-^1j_+^1\psi=(j_+^2j_-^1)',
\ey
\bay\label{ppt} 
j_-^2j^2_-\psib + j_-^1j_-^1\psi=(j_-^1j_-^2)'.
\ey

Therefore, using the scattering rule  
$$
\int\limits^{x}_{-\infty} \( j_-^2j^2_+\psib(y) + j_-^1j_+^1\psi(y)\) dy =j_+^2j_-^1 |_{-\infty}^{x}= j_+^2j_-^1 (x)-a(\l), 
$$
and 
$$
\int\limits^{x}_{-\infty} \( j_-^2j^2_-\psib(y) + j_-^1j_-^1\psi(y)\) dy= j_-^1j_-^2(x).  
$$
Finally,  we obtain
$$
X_1\jo_-(x)= \frac{i\jo_-(x)}{a(\l)}(j_+^2j_-^1 (x)-a(\l)) - \frac{i\jo_+(x)}{a(\l)}j_-^1j_-^2(x). 
$$

From this formula using $\jo _-J\jo_+=a(\l)$ we have 
$$
X_1 j_-^1(x)=\frac{i}{a}\(j_-^1j_+^2j_-^1-j_+^1j_-^1j_-^2\)-ij_-^1= \frac{i}{a}j_-^1\, \jo_-J\jo_+ -ij^1_-=ij_-^1-ij_-^1=0, 
$$
and 
$$
X_2 \jo_-(x)= \frac{i}{a}\(j_-^2 j_+^2 j_-^1-j_+^2j_-^1j_-^2\)-ij_-^2=ij_-^2. 
$$

Now we give  the proof of  identities \ref{ppi} and \ref{ppt}. 
To prove the first identity we take the first equation of the auxiliary linear system for $\jo_-$ 
$$
{j_-^1}'=-\frac{i\l}{2}j^1_- +\psib j_-^2
$$
and the second equation written for $\jo_+$
$$
{j_+^2}'=\psi j_+^1+\frac{i\l}{2}j_+^2. 
$$
Multiplying the first equation on $j_+^2$ and the second equation on $j^1_-$  and taking their sum we obtain \ref{ppi}. 

To prove \ref{ppt} we write equations for both components of $\jo_-$ 
$$
{j_-^1}'=-\frac{i\l}{2}j^1_- +\psib j_-^2,
$$
$$
{j_-^2}'=\psi j_-^1+\frac{i\l}{2}j_-^2. 
$$
Multiplying the first equation on $j_-^2$ and the second equation on $j^1_-$  and taking their sum we obtain \ref{ppt}.
We are done. 

The second set of formulas  for the action of $X_1$ on $\jo_+(x)$ can be proved along these lines. 
\qed

Now we are ready to prove  Theorem \ref{divfl}.  

\noi
{\it Proof.}   Since $Xa(\l)=0$ we have
\bey
X_1\xi(x,\l)&=& X_1\Im \log \js_-\js_+(x,\l)\\
&=& \frac{1}{2i}\(\frac{X_1\js_-}{\js_-}-\frac{X_1\bar{\js}_-}{\bar{\js}_-}\) +
\frac{1}{2i}\(\frac{X_1\js_+}{\js_+}-\frac{X_1\bar{\js}_+}{\bar{\js}_+}\).
\eey
Using the result of  Lemma \ref{hhg}  for the first bracket we have
\bey
\therefore\;=\frac{1}{2}\(\frac{-j_-^2\bar{j}^1_--|j_-^2|^2-\bar{j}_-^2j_-^1-|j_-^2|^2}{|\js_-|^2}\).
\eey 
Using the identity\footnote{This identity is obtained using Remark \ref{ffgr}.} 
\bay\label{zut}
|j_-^1|^2-|j_-^2|^2=\jo_-^T J \hat{\jo}_-=\f_{\leftarrow}^T J \hat{\f}_{\leftarrow}=1
\ey 
we  obtain
\bey
\therefore\;=\frac{-|\js_-|^2+1}{2\, |\js_-|^2}.
\eey
For the second bracket we have
\bey
\therefore\;=\frac{1}{2}\(\frac{j_+^1\bar{j}^2_++|j_+^1|^2+\bar{j}_+^1 j_+^2+|j_+^1|^2}{|\js_+|^2}\),
\eey 
and using the identity 
\bay\label{mir}
|j_+^1|^2-|j_+^2|^2=\jo_+^T J \hat{\jo}_+=\f_{\rightarrow}^T J \hat{\f}_{\rightarrow}=-1
\ey
we  obtain
\bey
\therefore\;=\frac{|\js_+|^2-1}{2\, |\js_+|^2}.
\eey
Taking the sum we   obtain
$$
\frac{1}{2}\(\frac{1}{|\js_-|^2} -\frac{1}{|\js_+|^2}\).
$$

The formulas  
$$
|a|e^{\xit +\c}=|j_+|^2,\qquad\qquad\qquad
|a|e^{\xit-\c}=|j_-|^2;
$$
follow from the definitions of $\Pi$ and $\U$ allow to put the computation into the final form. 
\qed

\begin{thm} \label{divfll}
The action of the second vector field $X_2=\{\bullet, \HH_2\}$ on the S-divisor  is given by the formula
\beq
X_2\xi(x,\l)=[-\l + 2\Im \psi(x)]X_1\xi(x,\l).
\eeq
\end{thm}
Note that  up to the sign the relation between the first and the second NLS flows is  the same as in the periodic case, \cite{MV}. 
We start with the analog of Lemma \ref{hhg}. This result is proved without computation of the gradients of  Jost solutions. 

\begin{lem}\label{ssfsd}
The action of the second vector field $X_2=\{\bullet, \HH_2\}$ on the Jost solutions is given by the formulas
\bey
X_2 \jo_+(x) &=& \jo_+'(x),\qquad\qquad\qquad\qquad X_2\jo_-(x)=\jo_-'(x).\\
\eey
\end{lem}
\noi
{\it Proof.}  Suppose we are interested   in the deformation of the operator $\partial_x-V(x,\l)$ defined by the formula 
$$
\partial_\tau V - \partial_x U + \[U,V\]=0.
$$
where the matrix $U=U(x,\l)$ and $\tau$ is the parameter corresponding to the deformation. How the solution $\jo(x,\l)$ of the problem 
$$
\(\partial_x-V\)\jo=0
$$
changes under such deformation? Differentiating this formula with respect to $\tau$ and substituting the expression  for $\partial_\tau V$  
we have after simple algebra 
$$
\(\partial_x-V\)\partial_\tau \jo=\(\partial_x-V\)U\jo.
$$
This implies 
$$
\partial_\tau \jo=U\jo+ \alpha \jo_-+\beta \jo_+,
$$
with some constants $\alpha=\alpha(\l)$ and $\beta=\beta(\l)$. In the case of the Jost solution $\jo=\jo_{\pm}$ the constants can be determined from the 
asymptotic behavior when $x\rightarrow \pm \infty$.  

Let us demonstrate first how these arguments produce the first result of Lemma \ref{hhg}. In this case 
$$
\partial_\tau \jo_+=\frac{i}{2}\sigma_3\jo_++ \alpha \jo_-+\beta \jo_+.
$$ 
Now letting $x\rightarrow +\infty$ and using the scattering rule we have 
$$
0=\partial_\tau \f_\rightarrow=\frac{i}{2}\sigma_3  \f_\rightarrow    + \alpha ( a \f_\leftarrow  +   b  \f_\rightarrow     ) +\beta \f_\rightarrow.
$$
Now  we  see that $\alpha=0$ and $\beta=\frac{i}{2}$. Therefore, 
$$
X_1\jo_+=\partial_\tau \jo_+= \frac{i}{2}(\sigma_3+I)\jo_+.
$$

In the case of the second flow the result follows in  a similar way. Let us prove it for $\jo_+$. The parameter $\tau$ consides with $x$ and $U=V$. In this case
$$
\partial_x \jo_+=\(- \frac{i \l }{ 2} \sigma_3 +Y_0\)\jo_++ \alpha \jo_-+\beta \jo_+.
$$
Now letting $x\rightarrow +\infty$ and using the scattering rule we have 
$$
 \frac{i \l }{ 2}\f_\rightarrow=-\frac{i\l}{2}\sigma_3  \f_\rightarrow    + \alpha ( a \f_\leftarrow  +   b  \f_\rightarrow     ) +\beta \f_\rightarrow.
$$
This implies that $\alpha=\beta=0$ and 
$$
X_2\jo_+=\partial_\tau \jo_+= \jo_+'.
$$
\qed

Now we are ready to prove  Theorem \ref{divfll}.  

\noi
{\it Proof.}   Since $Xa(\l)=0$ we have using the result of  Lemma \ref{ssfsd}  
\bey
X_2\xi(x,\l)&=& X_2\Im \log \js_-\js_+(x,\l)\\
&=& \frac{1}{2i}\(\frac{X_2\js_-}{\js_-}-\frac{X_2\bar{\js}_-}{\bar{\js}_-}\) +
\frac{1}{2i}\(\frac{X_2\js_+}{\js_+}-\frac{X_2\bar{\js}_+}{\bar{\js}_+}\)\\
&=& \frac{1}{2i}\(\frac{\js_-'}{\js_-}-\frac{\bar{\js}_-'}{\bar{\js}_-}\) +
\frac{1}{2i}\(\frac{\js_+'}{\js_+}-\frac{\bar{\js}_+'}{\bar{\js}_+}\)\\
&=& \frac{1}{2i}\frac{{\js_-'}{\bar{\js}_-}-{\bar{\js}_-'}{{\js}_-}}{|\js_-|^2} +
\frac{1}{2i}\frac{{\js_+'}{\bar{\js}_+}-{\bar{\js}_+'}{{\js}_+}}{|\js_+|^2}.
\eey

To compute the numerators we establish the  identity: 
\bay\label{vii}
{\js'}{\bar{\js}}-{\bar{\js}'}{{\js}}=(-i\l+2i \Im \psi(x)) \(|j^1|^2-|j^2|^2\),  \qquad \qquad \js=\js_{\pm}(x,\l). 
\ey
The original system 
$$
{j^1}'=-\frac{i\l}{2}j^1 +\psib j^2,
$$
$$
{j^2}'=\psi j^1 +\frac{i\l}{2}j^2;
$$
implies
$$
\js'=-\frac{i\l}{2}(j^1-j^2)+\psib j^2+\psi j^1.
$$ 
 After simple computaions  we obtain the stated identity
\bey
{\js'}{\bar{\js}}-{\bar{\js}'}{{\js}}&=&2i \Im \psi(x) \(|j^1|^2-|j^2|^2\)-\frac{i\l}{2} \[ (j^1-j^2)(\bar{j}^1+\bar{j}^2) +(\bar{j}^1-\bar{j}^2)({j^1}+{j^2})\]\\
&=&(-i\l+2i \Im \psi(x)) \(|j^1|^2-|j^2|^2\).  
\eey

Using \ref{vii}, \ref{zut} and \ref{mir},  we have
\bey
\therefore\;=(-\l+2 \Im \psi(x))\times \frac{1}{2}\(\frac{1}{|\js_-|^2} -\frac{1}{|\js_+|^2}\).
\eey
The second multiple is nothing but the action of the first flow on the divisor. 

\qed


\newpage

\section{ The  Poisson bracket.}
\subsection{The Poisson bracket for the $\Pi$ function.} Surprisingly, the bracket for the function $\Pi(x,\l)$ corresponding  two  different values of the spectral parameter can 
be expressed in a closed form.  We need to introduce some notations.  Let $\l,\mu$ in the upper half--plane
$$
\jo_+=\jo_+^{(2)}(x,\l),\qquad\qquad\qquad \jo_-=\jo_{-}^{(1)}(x,\l), 
$$ 
and 
$$
\gb_+=\gb_+^{(2)}(x,\m),\qquad\qquad\qquad \gb_-=\gb_{-}^{(1)}(x,\m). 
$$
Denote by 
$$
W\[\jo_{\pm},\gb_{\pm}\]=\jo_{\pm}(x,\l)J\gb_{\pm}(x,\mu)
$$
the Wronskian of two solutions. 
\begin{thm} \label{pbpi} For $\Pi(\l)=\Pi(x,\l),\;\U(\m)=\U(x,\m),$ and $\l,\; \mu$ in the upper half--plane  we have   
\bey
\{\Pi(\l),\Pi(\mu)\}& =  &
 \frac{2\Pi(\l)\Pi(\mu)}{a(\l)\, a(\mu)(\mu-\l)}\times \phantom{pppppppppppppppppppppppppppppppppppp} \\
&\times& \(\U(\l)\U(\mu)W^2\[\jo_-,\gb_-\](x)-\U^{-1}(\l)\U^{-1}(\mu)W^2[\jo_+,\gb_+](x)\).
\eey
\end{thm}
\noi
{\it Proof.}  \bey
\{\Pi(\l),\Pi(\mu)\}=  \{\frac{\js_+ \js_-}{a(\l)},\frac{\gs_+ \gs_-}{a(\m)}\} 
& =  & \{\js_+ ,\gs_+\}\frac{\js_- \gs_-}{a(\l)\,a(\m)} \\
& +  & \{\js_- ,\gs_-\}\frac{\js_+ \gs_+}{a(\l)\,a(\m)}  \\
& +  & \{\js_+ \js_-,\frac{1}{a(\m)}\}\frac{\gs_+ \gs_-}{a(\l)} \\
& +  & \{\frac{1}{a(\l)},\gs_+ \gs_-\}\frac{\js_+ \js_-}{a(\m)}. \\
\eey

\noi
{\it Step 1.} Using Lemma   \ref{vjs} one can compute for the first term 
\bey
\{\js_+ ,\gs_+\}\frac{\js_- \gs_-}{a(\l)\,a(\m)}=\frac{2}{a^2(\l) a^2(\m) (\m -\l)}\times    
[&-&\js_+ \js_- \gs_+ \gs_- \(\jo_+^T J \gb_+\)\(\jo_-^T J \gb_-\) \\ 
&+&\js_+ \js_- \gs_- \gs_- \(\jo_+^T J \gb_+\)\(\jo_-^T J \gb_+\)  \\
&+& \js_- \js_- \gs_+ \gs_- \(\jo_+^T J \gb_+\)\(\jo_+^T J \gb_-\)  \\
 &-& \js_- \js_- \gs_- \gs_- \(\jo_+^T J \gb_+\)\(\jo_+^T J \gb_+\)].  \\
\eey

\noi
{\it Step 2.} Similar one can compute for the second term 
\bey
\{\js_- ,\gs_-\}\frac{\js_+ \gs_+}{a(\l)\,a(\m)}=\frac{2}{a^2(\l) a^2(\m) (\m -\l)}\times    
[&+&\js_+ \js_- \gs_+ \gs_- \(\jo_+^T J \gb_+\)\(\jo_-^T J \gb_-\) \\ 
&-&\js_+ \js_- \gs_+ \gs_+ \(\jo_+^T J \gb_-\)\(\jo_-^T J \gb_-\)  \\
&-& \js_+ \js_+ \gs_+ \gs_- \(\jo_-^T J \gb_-\)\(\jo_-^T J \gb_+\)  \\
&+& \js_+ \js_+ \gs_+ \gs_+ \(\jo_-^T J \gb_-\)\(\jo_-^T J \gb_-\)].  \\
 \eey

\noi
{\it Step 3.}          For the third term 
\bey
\{\js_+ \js_-,\frac{1}{a(\m)}\}\frac{\gs_+ \gs_-}{a(\l)}=\frac{2}{a^2(\l) a^2(\m) (\m -\l)}\times    
[&+&\js_+ \js_+ \gs_+ \gs_- \(\jo_-^T J \gb_-\)\(\jo_-^T J \gb_+\) \\ 
&-& \js_- \js_- \gs_+ \gs_- \(\jo_+^T J \gb_+\)\(\jo_+^T J \gb_-\)].  \\
\eey

\noi
{\it Step 4.}          For the fourth  term 
\bey
\{\frac{1}{a(\l)},\gs_+ \gs_-\}\frac{\js_+ \js_-}{a(\m)}=\frac{2}{a^2(\l) a^2(\m) (\m -\l)}\times    
[&+&\js_- \js_+ \gs_+ \gs_+ \(\jo_-^T J \gb_-\)\(\jo_+^T J \gb_-\) \\ 
&-& \js_- \js_+ \gs_- \gs_- \(\jo_+^T J \gb_+\)\(\jo_-^T J \gb_+\)].  \\
\eey

\noi
{\it Step 5.} Taking sum of the results of the previous four steps after cancellations we have 
\bey
\{\Pi(\l),\Pi(\mu)\}= \frac{2}{a^2(\l) a^2(\m) (\m -\l)}\times [ \js_+^2  \gs_+^2  \(\jo_-^T J \gb_-\)^2 
- \js_-^2 \gb_-^2 \(\jo_+^T J \gb_+\)^2].
\eey 
After simple algebra we obtain the formula stated above.

\qed

\

\end{document}